\documentclass[%
reprint,
superscriptaddress,
nofootinbib,
amsmath,amssymb,
aps,
pra,
longbibliography,
]{revtex4-2}

\usepackage{graphicx}
\usepackage{dcolumn}
\usepackage{bm}

\usepackage{hyperref}
\hypersetup{
    colorlinks=true,
    linkcolor=blue,
    filecolor=blue,      
    urlcolor=blue,	
    citecolor=blue
}
\usepackage{physics}
\usepackage{braket}
\usepackage{siunitx}
\usepackage{booktabs}
\usepackage{qcircuit}
\usepackage{floatrow}
\usepackage[shortlabels]{enumitem}
\usepackage{subcaption}
\usepackage{xcolor}

\usepackage[version=4]{mhchem}
\usepackage{upgreek,ulem}

\definecolor{hughcol}{HTML}{A1246B}

\newcommand{\al}{\alpha}
\newcommand{\be}{\beta}

\newcommand{\RHF}{\Phi_{\mathrm{RHF}}}
\newcommand{\sigg}{\upsigma_{\text{g}}}
\newcommand{\sigu}{\upsigma_{\text{u}}}
\newcommand{\bsigg}{\bar{\upsigma}_{\text{g}}}

\newcommand{\upo}{\text{o}}
\newcommand{\uph}{\text{h}}
\newcommand{\upa}{\text{a}}
\newcommand{\upb}{\text{b}}
\newcommand{\upe}{\text{e}}
\newcommand{\upx}{\text{x}}
\newcommand{\upy}{\text{y}}
\newcommand{\upz}{\text{z}}
\newcommand{\ups}{\text{s}}
\newcommand{\upp}{\text{p}}

\newcommand{\bups}{\bar{\text{s}}}

\newcommand{\ox}{Physical and Theoretical Chemistry Laboratory, University of Oxford, South Parks Road, Oxford, OX1 3QZ, UK}
\newcommand{\cam}{Yusuf Hamied Department of Chemistry, University of Cambridge, Lensfield Road, Cambridge, CB2 1EW, UK}

\UseRawInputEncoding
\begin{document}
	

\title{Spin-coupled molecular orbitals: chemical intuition meets quantum chemistry}

\author{Daniel Marti-Dafcik}
\email{dmartidafcik@gmail.com}
\affiliation{\ox}
\author{Nicholas Lee}
\affiliation{\ox}
\author{Hugh G.~A. Burton}
\affiliation{\ox}
\affiliation{\cam}
\author{David P. Tew}
\email{david.tew@chem.ox.ac.uk}
\affiliation{\ox}

\date{\today}

\begin{abstract}

Molecular orbital theory is powerful both as a conceptual tool
for understanding chemical bonding, and as a theoretical framework for ab initio quantum chemistry.
Despite its undoubted success, MO theory has well documented shortcomings, most notably that it 
fails to correctly describe diradical states and homolytic bond fission. In this
contribution, we introduce a generalised MO theory that includes spin-coupled radical states.
We show through archetypical examples that when bonds break, the electronic state transitions between a 
small number of valence configurations, characterised by occupation of both delocalised molecular orbitals 
and spin-coupled localised orbitals.
Our theory provides a model for chemical bonding that is both chemically intuitive and qualitatively
accurate when combined with ab initio theory. Although exploitation of our
theory presents significant challenges for classical computing, the predictable structure of
spin-coupled states is ideally suited to algorithms that exploit quantum computers. Our
approach provides a systematic route to overcoming the initial state overlap problem and unlocking 
the potential of quantum computational chemistry.
\end{abstract}

\maketitle
\raggedbottom

\section{Introduction }\label{sec:intro}

Solving Schr\"odinger's equation to characterise the electronic structure of molecules and compute the
Born--Oppenheimer potential energy surface remains a central goal of quantum chemistry. Modern electronic
structure theory has arisen from nearly a century of sustained research and development, and the vast majority of modern
methods are based on Molecular Orbital (MO) theory,\cite{Mulliken1928I, Mulliken1928II, Lennard-Jones1949, Hall1950} 
either in the form of Kohn--Sham Density Functional Theory\cite{Kohn1965} 
or post-Hartree--Fock correlation methods.\cite{HelgakerBook} 
These approaches came to prominence in no small part due to the natural mapping of the MO representation 
of the many-electron wavefunction to linear algebra and matrix operations that can be performed with
high efficiency on classical digital computers.\cite{Roothaan1951,Slater1953}

Despite the success of MO theory, it has well-documented shortcomings.\cite{Lykos1963,Tew2007,Slater1932,VanVleck1935} Closed-shell ground states
are typically well approximated by a single configuration of electrons in spin-orbitals, but
open-shell states, or cases with many competing low-energy configurations, are not. In practice, the accurate description of
fundamental chemical processes, such as bond breaking using MO theory, requires a very large number of excited
configurations,\cite{Roos1980a, Roos1980b} and sophisticated optimisation techniques.\cite{White1992, White1993, White1999, Huron1973}
As a consequence, any connection between simple chemical models and data from high-level ab initio calculations is obfuscated.

Looking to the future, it is increasingly probable that quantum computers will become available for
scientific computing.\cite{Reiher2017, Arute2019, AIQuantum2020, Huggins2022b, Bluvstein2023}
Many of the constraints on the way that many-electron wavefunctions
are represented and optimised in modern MO-based theories do not apply to quantum algorithms and
we were therefore motivated to examine the electronic structure of bond formation and breaking afresh.

In this article, we introduce a generalised MO theory that includes spin-coupled radical states. Our theory provides
a chemically intuitive picture of bonding that directly maps to a highly compact representation of the
many-body wavefunction. 
Contrary to the current consensus among the electronic structure community, we find that 
it is not necessary to use sophisticated multi-reference expansions with very large numbers of variational
parameters to describe bond breaking. Instead, we show that the electronic state undergoes transitions between
a small number of valence states, each with an easily understood spin-coupled electronic configuration. 

Furthermore, we show that our representation encodes strong correlation through prescribed patterns of entanglement, which
is ideally suited to quantum computing. We demonstrate that our approach provides
a powerful route to construct the sufficiently accurate reference states that are required for the practical application of 
fault-tolerant algorithms, such as quantum phase estimation\cite{Aspuru-Guzik2005, Kitaev1995, Abrams1999} and related approaches.\cite{Lin2020a, Lin2022, Dong2022, Wang2023, Wang2023b}


\section{Molecular hydrogen}
\label{sec:h2}

The prototypical chemical bond between two hydrogen atoms
exemplifies the language and conceptual framework within which electronic structure theories operate.
In MO theory, the left $L$ and right $R$ $1\ups$ atomic orbitals combine to form bonding
$\sigg$ and anti-bonding $\sigu$ MOs that transform as the irreducible representations of the $\mathrm{D_{\infty h}}$
molecular point group.\cite{Lennard-Jones1949, Hall1950} The wavefunction is written as the closed-shell singlet, doubly
occupying the bonding MO, which is the Slater determinant used in Hartree--Fock theory
$\ket{\Phi_{\mathrm{RHF}}} = \ket{\sigg \bsigg}$.
The MO description is accurate around the equilibrium bond length, but deteriorates as the molecule dissociates.
This well-known failure arises from the inclusion of ionic terms in the wavefunction, which results from
delocalised nature of the MOs\cite{Slater1953}
\begin{equation} 
\ket{\RHF} = \frac{1}{2}\Big( \underbrace{\ket{\ups_L\bups_L}  + \ket{\ups_R\bups_R}}_\text{ionic} + \underbrace{\ket{\ups_L\bups_R} - \ket{\bups_L\ups_R}}_{\text{covalent}} \Big). 
\end{equation}
The strong correlation that localises the electrons on opposite atoms is absent in the MO picture. In
MO-based theories, this strong correlation is introduced through mixing with excited configurations, and for
\ce{H2} the correct behaviour for the binding curve is obtained using two configurations
\begin{equation}
        \ket{\Psi} = c_1 \ket{\sigg\bar{\upsigma}_\text{g}} + c_2 \ket{\sigu\bar{\upsigma}_\text{u}}.
\end{equation}
In the general case, the number of excited configurations required to recover the 
correct electronic structure grows exponentially with the number of localised electrons.
Modern MO-based methods for strong correlation therefore typically employ
sophisticated wavefunction representations to optimise large numbers of variational parameters.

An alternative representation that predates MO theory is the valence bond (VB) approach of Heitler and London.\cite{Heitler1927} The VB
wavefunction is the open-shell singlet spin-coupled state of $1\ups$ electrons on each \ce{H} atom 
\begin{equation}
        \ket{\Phi_\mathrm{VB}} = \mathcal{N} \qty(\ket{\ups_L\bups_R} - \ket{\bups_L\ups_R}).
\end{equation}
where the normalisation constant $\mathcal{N} = (2+2S_{LR}^2)^{-1/2}$ depends on the orbital overlap $S_{LR} = \braket{\ups_L | \ups_R}$.
The VB description also captures the bonding interaction, which arises through electron
delocalisation resulting from the overlap of the $L$ and $R$ $1\ups$ orbitals.
Approaches based on VB wavefunctions, however, are poorly suited to both classical and quantum computing due to
reliance on non-orthogonal orbitals, which introduces cumbersome overlap terms in the working equations.\cite{Gerratt1971,Goddard1973,Li1994}

Notwithstanding the differences in the MO and VB representations, the
physical process of bond breaking is clear: the electronic state transitions from a
singlet spin-coupled bonding $\sigg^2$ configuration in the equilibrium region to a singlet spin-coupled
diradical $\ups_L^1 \ups_R^1$ configuration at dissociation. On a classical digital computer, the
basis transformations required to combine configurations of different orbitals are costly
and often impractical. On a digital quantum computer, basis transformations pose no difficulty. 
Orbital rotations are encoded through products of exponentials of single excitation and de-excitation operators,\cite{Kivlichan2018} 
and the corresponding quantum circuits have linear depth in orbital number and minimal qubit connectivity.\cite{AIQuantum2020}

We therefore choose to generalise MO theory by expressing the wavefunction as a superposition of
configurations where electrons occupy both delocalised MOs and localised AOs. This representation
directly encodes both the covalent interactions responsible for bonding and the
the strong correlation that localises electrons at               
different sites, giving a physically accurate trial wavefunction across the whole binding curve 
without requiring large numbers of variational parameters.

\begin{figure}[t!]
	\centering
	\includegraphics[width=1\textwidth]{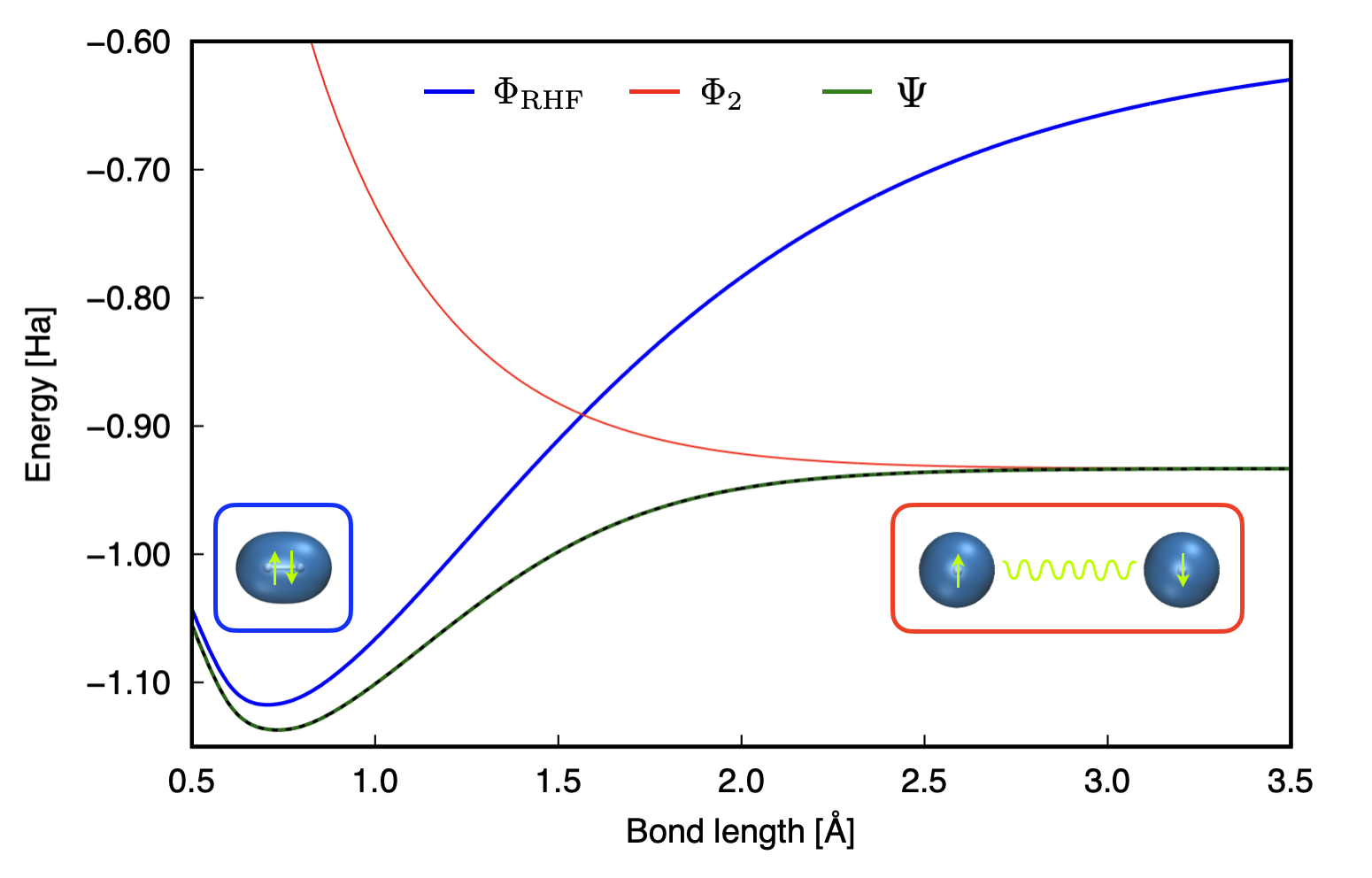}
	\caption{Dissociation of \ce{H2} in the STO-3G basis using generalised spin-coupled orbitals.}
	\label{fig:h2_fig}
\end{figure}

For \ce{H2}, the singlet coupled $\ups_L \ups_R$ diradical state is
\begin{equation}
	\ket{\Phi_2} = \frac{1}{\sqrt{2}}\qty(\ket{\ups_L\bups_R} - \ket{\bups_L\ups_R}) = \frac{1}{\sqrt{2}}\qty(\ket{\sigg\bar{\upsigma}_\text{g}} - \ket{\sigu\bar{\upsigma}_\text{u}}).
\end{equation}
where now $\ups_L = {1 \over \sqrt2} ( \sigg + \sigu)$ and $\ups_R = {1 \over \sqrt2} ( \sigg - \sigu)$ are orthogonal
orbitals localised on the left and right atoms. In contrast
to the Heitler--London wavefunction, this is a purely repulsive state.\cite{Malrieu2008} The correct binding curve (Figure \ref{fig:h2_fig})
is obtained through the linear combination of the $\sigg^2$ and $\ups_L\ups_R$ states
\begin{equation}
        \ket{\Psi} = c_1 \ket{\Phi_{\mathrm{RHF}}} + c_2 \ket{\Phi_2}. 
\end{equation}

\noindent The $\ket{\Phi_{\mathrm{RHF}}}$ state is uncorrelated, since 
a single Slater determinant is an antisymmetrised orbital product corresponding to
an independent electron wavefunction.  $\ket{\Phi_2}$ is an entangled, or strongly correlated state, 
since it is a linear combination of determinants that cannot be converted to a single determinant through orbital rotation. 
Both states have low complexity since their structures are determined by symmetry considerations.
The weights of the entanglement structure in $\ket{\Phi_2}$ are specified through the rules of spin angular 
momentum coupling to obtain a singlet spin molecular ground state. 

In this representation, the bonding interaction arises through increasing character of the
delocalised $\sigg^2$ configuration, which is driven by the favourable kinetic energy of the delocalised state
and the attractive electron-nucleus Coulomb interaction.
Conversely, the dominant electron correlation process for the H$_2$ bond is the
localisation of the electrons on opposite atoms, introducing partial diradical character,
and this correlation increases in strength as the bond length increases.
This transition from delocalised to localised electronic states is equivalent to the metal-insulator transition in the 
Fermi--Hubbard model, where the \ce{H2} bond length plays a similar role to the ratio $U/t$ between the on-site 
repulsion $U$ and the hopping term $t$. 

In the following, we demonstrate how these simple concepts can be generalised to more complex cases
where multiple bonds are broken. We show that the strongly correlated state at 
dissociation is highly structured and that the structure is determined solely by symmetry and spin-angular momentum
coupling considerations. Furthermore, we show that the transition from the simple covalent equilibrium 
bonding structure to the strongly correlated dissociated state is represented by a small number of
intermediate configurations, each with their own spin-coupled structure. We first summarise
the configuration state functions that we use to represent spin-coupled open-shell configurations.

\section{Configuration State Functions}\label{sec:csfs}

The exact molecular wavefunction is an eigenstate of both the total spin angular momentum operator $\hat{S}^2$ and
the spin projection operator $\hat{S}_z$ with quantum numbers $S,M$.
A single Slater determinant, the many-electron basis function in MO-based theories,
is an eigenfunction of $\hat{S}_z$, but not of $\hat{S}^2$, except for
closed-shell and high-spin open-shell configurations. The physically relevant functions for describing
open-shell states with localised orbitals are configuration state functions (CSF), which
are eigenstates of $\hat{S}^2$ with quantum numbers $S,M$. 

A CSF is defined by a specification of the open-shell orbitals and a specification of the spin-coupling pattern,
which is determined by the order and manner in which the spin-$\frac{1}{2}$ particles in the open-shell orbitals are coupled.
Formally, our CSFs take the form
\begin{equation}
        \ket{\Phi} = \mathcal{N} \hat{\mathcal{A}} \qty(\ket{\phi_c \bar{\phi}_c \dots} \mathcal{O}_{SM}^{N,i}(\phi_o \bar{\phi}_o \dots)  ).
\label{eq:csf}
\end{equation}
where $\hat{\mathcal{A}}$ is the antisymmetric permutation operator,
$c$ denotes the doubly occupied closed-shell orbitals, $o$ are the open-shell orbitals, and
$\mathcal{O}_{SM}^{N,i}(\phi_o \bar{\phi}_o \dots)$ is an open-shell state with $N$ electrons in orbitals $\{\phi_o,\bar{\phi}_o\}$
coupled with spin quantum numbers $S$ and $M$ in a spin-coupling pattern labelled by $i$, and $\mathcal{N}$ is a normalisation constant. 
All occupied spin-orbitals are mutually orthogonal and normalised. 

The open-shell state $\mathcal{O}_{SM}^{N,i}$ for coupled angular momenta can be expressed in terms of the uncoupled
Slater determinant representation of $N$ electrons in open-shell orbitals $\{\phi_o,\bar{\phi}_o\}$
through the Clebsch--Gordan coefficients for the spin-coupling pattern. A list of relevant CSF wavefunctions is
provided in Appendix \ref{sec:spin_eigenfunctions}. 

To streamline presentation, we use a simplified notation for $\ket{\Phi}$ in Eq \ref{eq:csf}, 
listing only the occupancy of the spatial orbitals and suppressing $\hat{\mathcal{A}}$ and $\mathcal{N}$:
\begin{equation}
        \ket{\Phi} = \ket{\phi_c^2 \dots} \mathcal{O}_{SM}^{N,i}(\phi_o \dots) .
\end{equation}

Classical digital approaches that use a CSF basis rather than a Slater determinant basis, such as spin-adapted FCI quantum Monte Carlo,\cite{LiManni2020}
are constrained to use one set of orbitals and one coupling scheme with an orthonormal basis of CSFs.
This restriction allows Hamiltonian matrix elements to be evaluated using efficient group theoretical approaches without explicitly building the Slater determinant expansion of each CSF, which has exponential cost.\cite{Paldus1974,Shavitt1977}
Quantum algorithms are not constrained in this way and we are free to use orbitals and coupling patterns
that directly encode the physical interactions.
Furthermore, we can combine AO configurations for coupled high-spin 
atomic fragments with MO configurations for delocalised covalent bonds, as well as configurations 
intermediate between the these extremes. 

The physical interactions follow established rules connected to 
the relative strengths of exchange and delocalisation energies, and the 
CSFs required to represent the electronic state compactly are straightforwardly constructed. For example,
the electron configurations of dissociated atomic fragments follow Hund's rules, where electron spins couple to form
a local high-spin state, and the net electron spin angular momenta of the atoms
couple to form low-spin states. Bonding interactions in the molecule are formed when paired electrons occupy
delocalised molecular orbitals.
We now illustrate this framework using examples that pose significant challenges for
traditional approaches.

\section{Homolytic bond cleavage}\label{sec:bond}

\subsection{Breaking a triple bond: \ce{N2}}\label{sec:n2}

\begin{figure*}
\begin{minipage}{0.61\columnwidth}
\begin{subfigure}[t]{\textwidth}
		\centering
		\includegraphics[width=\textwidth]{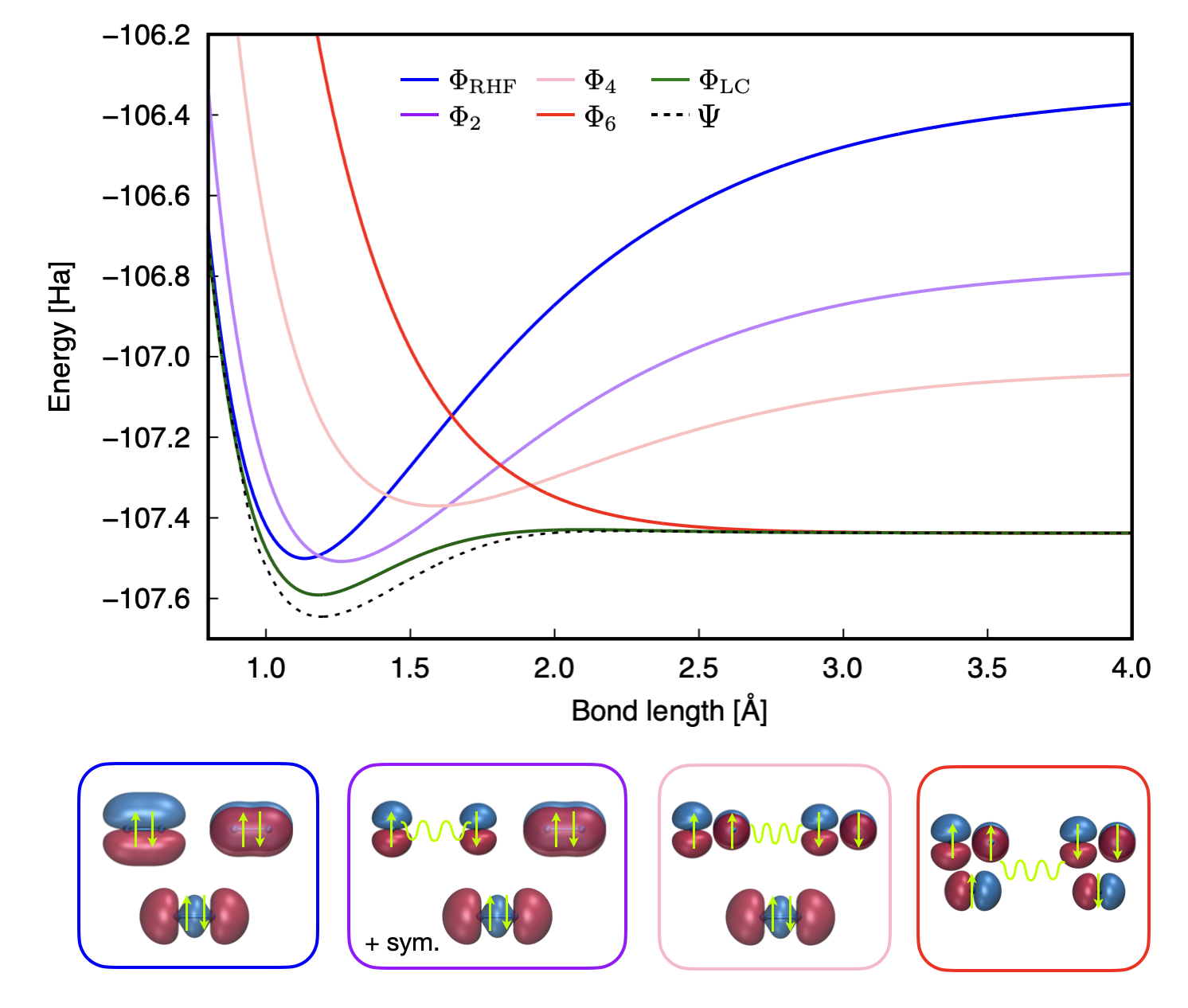}
		\caption{Energies and configurations of the valence CSFs.}
		\label{fig:n2_mainfig}
\end{subfigure}
\end{minipage}
\begin{minipage}{0.38\columnwidth}
\begin{subfigure}[t]{\textwidth}
	\centering
	\includegraphics[width=\textwidth]{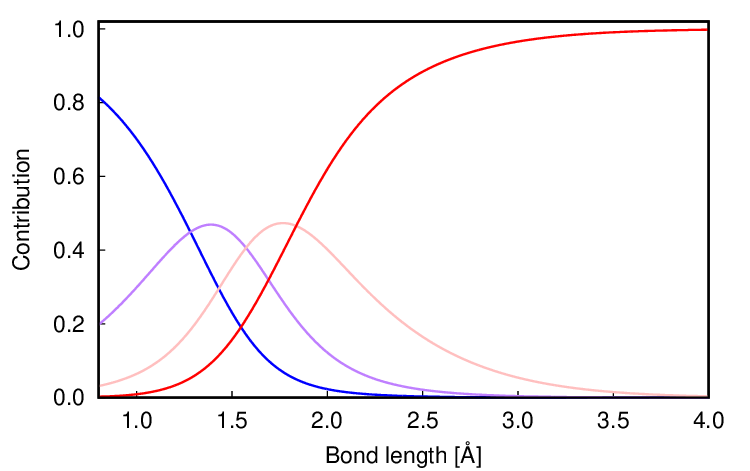}
	\caption{Coefficients of each CSF in $\Phi_\text{LC}$.}
	\label{fig:n2_csf_lin_comb}
\end{subfigure}
\begin{subfigure}{\textwidth}
	\centering
	\includegraphics[width=\textwidth]{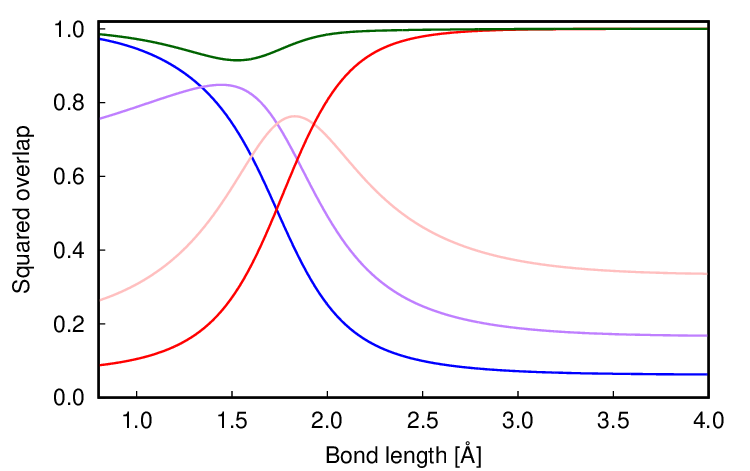}
	\caption{Squared overlaps $\abs{\braket{\Psi|\Phi_{\mathrm{i}}}}^2$.}
	\label{fig:n2_csf_fid}
\end{subfigure}
\end{minipage}
\centering
	\caption{The dissociation of \ce{N2} in a STO-3G basis in the six 2\upp\ valence space using generalised MO theory.  \label{fig:n2}}
\end{figure*}
Dissociation of the nitrogen molecule is a well-documented challenge for electronic structure theory\cite{Laidig1987,
Kowalski2000,Piecuch2001,Larsen2000,Chan2004,Bytautas2011,Parkhill2009,Gidofalvi2008,Small2009,Booth2009,Thom2010} 
because the six electrons in the triple bond become very strongly correlated when the bond elongates. Even when
only considering the six $2\upp$ valence orbitals\cite{VanVoorhis2001} there are 56 Slater determinants with the correct symmetry
that contribute to the wavefunction. However, using generalised MO theory, the wavefunction is recovered to better than
92\% across the whole binding curve with only four CSFs (Figure~\ref{fig:n2}).

MO theory predicts 
a $1\upsigma_\text{g}^2 1\upsigma_\text{u}^2 2\upsigma_\text{g}^2 2\upsigma_\text{u}^2 3\upsigma_\text{g}^2 1\uppi_{\text{u}}^4$
configuration for the \ce{N2} molecule, with a ${}^1\Sigma_\text{g}^+$ ground state. The corresponding CSF is simply
the closed-shell Hartree--Fock wavefunction 
\begin{align}
\ket{\RHF} &= \ket{ \mathcal{C} } \ket { 3\sigg^2 \, 1\uppi_{\text{u,x}}^2 \, 1\uppi_{\text{u,y}}^2 }
\end{align}
where $ \ket{\mathcal{C} } = \ket{ 1\upsigma_\text{g}^2 1\upsigma_\text{u}^2 2\upsigma_\text{g}^2 2\upsigma_\text{u}^2 }$ is 
the contribution from the core orbitals that are not involved in bonding.
At dissociation, each atom has a $1\ups^22\ups^22\upp^3$ electronic configuration with spins locally coupled to a ${}^4P$ on each atom,
which then couple through space to give a ${}^1\Sigma_\text{g}^+$ molecular state with the CSF structure
\begin{align}
\ket{\Phi_6} &= \ket{ \mathcal{C} } \mathcal{O}_{00}^{6,1}(\upx_L \upy_L \upz_L \upx_R \upy_R \upz_R )
\end{align}
where 
$\upz_L = \frac{1}{\sqrt 2} ( 3\upsigma_g + 3\upsigma_u )$, $\upz_R = \frac{1}{\sqrt 2} ( 3\upsigma_g - 3\upsigma_u )$, 
$\upx_L = \frac{1}{\sqrt 2} ( 1\uppi_{u,x} + 1\uppi_{g,x} )$, $\upx_R = \frac{1}{\sqrt 2} ( 1\uppi_{u,x} - 1\uppi_{g,x} )$ etc
are localised atomic orbitals.
The spin function $\mathcal{O}_{00}^{6,1}$ spin-couples the three-electron quartets on 
each atom into a singlet, and is a specific linear combination of 20 open-shell determinants [Eq.~\eqref{eqn:vcsf_n6}].
The closed-shell part $ \ket{\mathcal{C} }$ is invariant to rotations among its constituent orbitals making it equal to
$\ket{ 1\ups_L^2 1\ups_R^2 2\ups_L^2 2\ups_R^2 }$.
As the bond elongates, one $\uppi$ bond breaks first , then both $\uppi$ bonds, and finally the $\upsigma$ bond. The CSFs
that describe these intermediate states with one and two broken bonds are
\begin{align}
\ket{\Phi_4} &= \ket{ \mathcal{C} } \ket{ \upsigma_\text{g}^2 } \mathcal{O}^{4,1}_{00}(\upx_L \upy_L \upx_R \upy_R ) \\
\ket{\Phi_2} &= \ket{\Phi_{2x}} + \ket{\Phi_{2y}} \\
\ket{\Phi_{2x}} &= \ket{ \mathcal{C} } \ket{ \upsigma_\text{g}^2 \uppi_{\text{u,y}}^2 } \mathcal{O}^{2,1}_{00}(\upx_L \upx_R)  \\
\ket{\Phi_{2y}} &= \ket{ \mathcal{C} } \ket{ \upsigma_\text{g}^2 \uppi_{\text{u,x}}^2 } \mathcal{O}^{2,1}_{00}(\upy_L \upy_R)
\end{align}
where $\ket{\Phi_{2x}}$ and $\ket{\Phi_{2y}}$ are a degenerate pair.
The state $\mathcal{O}^{4,1}_{00}$ describes atomic triplet states coupled to form a singlet overall and
$\mathcal{O}^{2,1}_{00}$ describes atomic doublet states coupled to a singlet. The representations of
$\mathcal{O}^{6,1}_{00}$, $\mathcal{O}^{4,1}_{00}$ and $\mathcal{O}^{2,1}_{00}$ in terms of 
Slater determinants are collected in Appendix \ref{sec:spin_eigenfunctions}.
In this formulation, the localised orbitals $\upx_L,\upx_R,\upy_L,\upy_R,\upz_L,\upz_R$ are mutually orthogonal and are
therefore not true atomic orbitals except at infinite separation. 

\begin{figure*}[!t]
  \includegraphics[width=\textwidth]{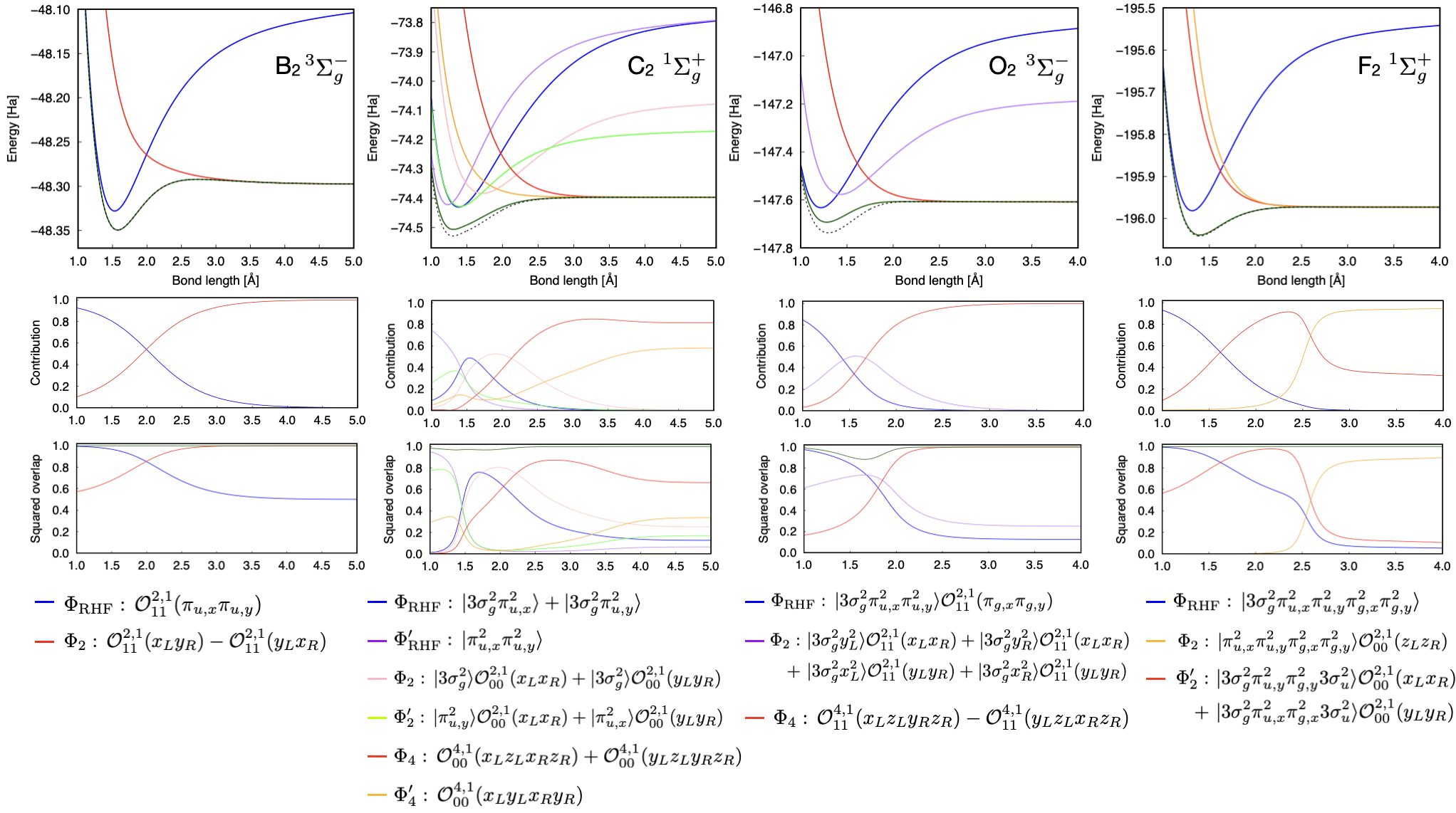}
  \caption{CSF energies, configurations and weights in the binding of first row diatomics.}
  \label{fig:diatomics}
\end{figure*}

Just as for H$_2$, the RHF state dissociates to the incorrect limit due to the presence of
unphysical ionic terms in the wavefunction, and the exact dissociated state is the CSF where the high-spin
atomic states couple to form a singlet (Figure~\ref{fig:n2_mainfig}). 
The CSFs with bond orders of two and one behave similarly to 
the RHF state but with higher energy in the bonding region, lower ionic destabilisation at dissociation,
and with minima at progressively longer bond lengths, entirely consistent with
their bonding character.
The closed-shell orbitals used in these CSFs were obtained from the RHF wavefunction, but very similar
curves are obtained if the orbitals of each CSF are relaxed, for example through RASSCF optimisation (restricted
active space self consistent field).\cite{Malmqvist1990}

Figure \ref{fig:n2_mainfig} also displays the energy of the exact ground state in the (6e,6o) active space of the
bonding electrons and the energy obtained from the variationally optimised linear combination of the four valence CSFs.
The optimised coefficients and the overlap of each state with the exact ground state are displayed in figures
\ref{fig:n2_csf_lin_comb} and \ref{fig:n2_csf_fid}.

While the optimised linear combination of these four CSFs is dominated by the RHF configuration at equilibrium,
electrons in a bond correlate by localising on opposite atoms, occupying low-energy spin-coupled states,
leading to non-negligible contributions
from the CSFs with lower bond orders and an equilibrium bond length longer than that of the RHF state.
Just as for H$_2$, the strength of these correlation
processes increases as the bond elongates. In this formulation, only four CSFs are required to 
obtain an accurate binding curve with an overlap with the exact state of at least $92\%$ at all bond lengths.
The largest deviations are found in the bonding region where there is significant additional dynamic correlation among
the six electrons in the triple bond. 

The state at dissociation is strongly correlated in the sense that the electrons
are localised on opposite atoms, and is a linear combination of 20 Slater determinants.
However, this state is described exactly by a single CSF. The complexity is low; the coefficients in the
determinantal representation are determined entirely by symmetry and spin angular momentum considerations.

Our generalised spin-coupled MO theory provides a highly intuitive description of chemical bonding that maps 
directly onto clearly identifiable contributions to the many-electron wavefunction. Since each contribution
is composed of different orbitals (localised and covalent) and different spin-coupling patterns, the states
are non-orthogonal and optimisation of the linear combination requires non-orthogonal configuration interaction (CI) theory.\cite{Fukutome1988,Ayala1998,Sundstrom2014,Thom2009,Burton2019c}
In the context of fault-tolerant quantum algorithms, such as quantum phase estimation, using $\ket{\Phi_\text{LC}}$ 
as the reference wavefunction instead of $\ket{\Phi_\text{RHF}}$ translates into more than an order of magnitude reduction in runtime at stretched geometries
(see Section \ref{sec:quantum_algos}).

\subsection{First row diatomics}\label{sec:diat}

Our spin-coupled generalised MO formalism is straightforwardly generalised to the bonding motifs of all of the first-row diatomic molecules, 
irrespective of whether the electronic ground state is singlet or triplet spin-coupled. Starting from the HF
state given by standard MO theory, CSFs with incrementally lower bond order are generated by 
localising the electrons in the weakest remaining bond into the corresponding non-bonding atomic orbitals. 
The spin coupling pattern follows Hund's rules, where
electrons first spin-couple with other unpaired electrons on the same atom to give high-spin atomic states,
which then spin-couple to give the appropriate overall spin state for the molecule. The $\mathrm{D_{\infty h}}$ molecular point group 
symmetry is satisfied by taking the symmetry adapted linear combinations generated by the symmetry operations.

In Figure \ref{fig:diatomics} we plot the potential energy curves for the valence CSFs and
the accuracy of the linear combination with respect to full CI in the ($N$e, 6o) active space for 
\ce{B2}, \ce{C2}, \ce{O2} and \ce{F2}. 
The CSFs are also listed. In all cases accurate binding curves and overlap with the FCI state of at least 88\% are obtained, with the
largest deviation for the bonding region of \ce{O2} where significant additional dynamic correlation contributes. The triplet state
\ce{B2} transitions from the $\uppi^2$ single-bond configuration to a triplet diradical state at dissociation.
For \ce{C2}, configurations with either $\upsigma$ or $\uppi$ occupation are both important since the  
large \text{sp} mixing, arising from the small $\text{s}$-$\text{p}$ orbital energy gap, destabilises the $\upsigma$ configurations. 
In the dissociation limit, the ground state tends towards the spin-coupled atomic states, where the electrons are evenly 
distributed among the $\text{p}_x$,$\text{p}_y$,$\text{p}_z$ orbitals. This is also the reason that \ce{F2} requires both the
$\upsigma$-hole and $\uppi$-hole configurations at dissociation. The triplet state \ce{O2} follows exactly the same
physical rules, but the three spin-coupled states contain multiple symmetry-related components.

\subsection{Breaking two single bonds: \ce{H2O}}\label{sec:h2o}
\begin{figure}[b]
	\begin{subfigure}{1\textwidth}
		\centering
		\includegraphics[width=\textwidth]{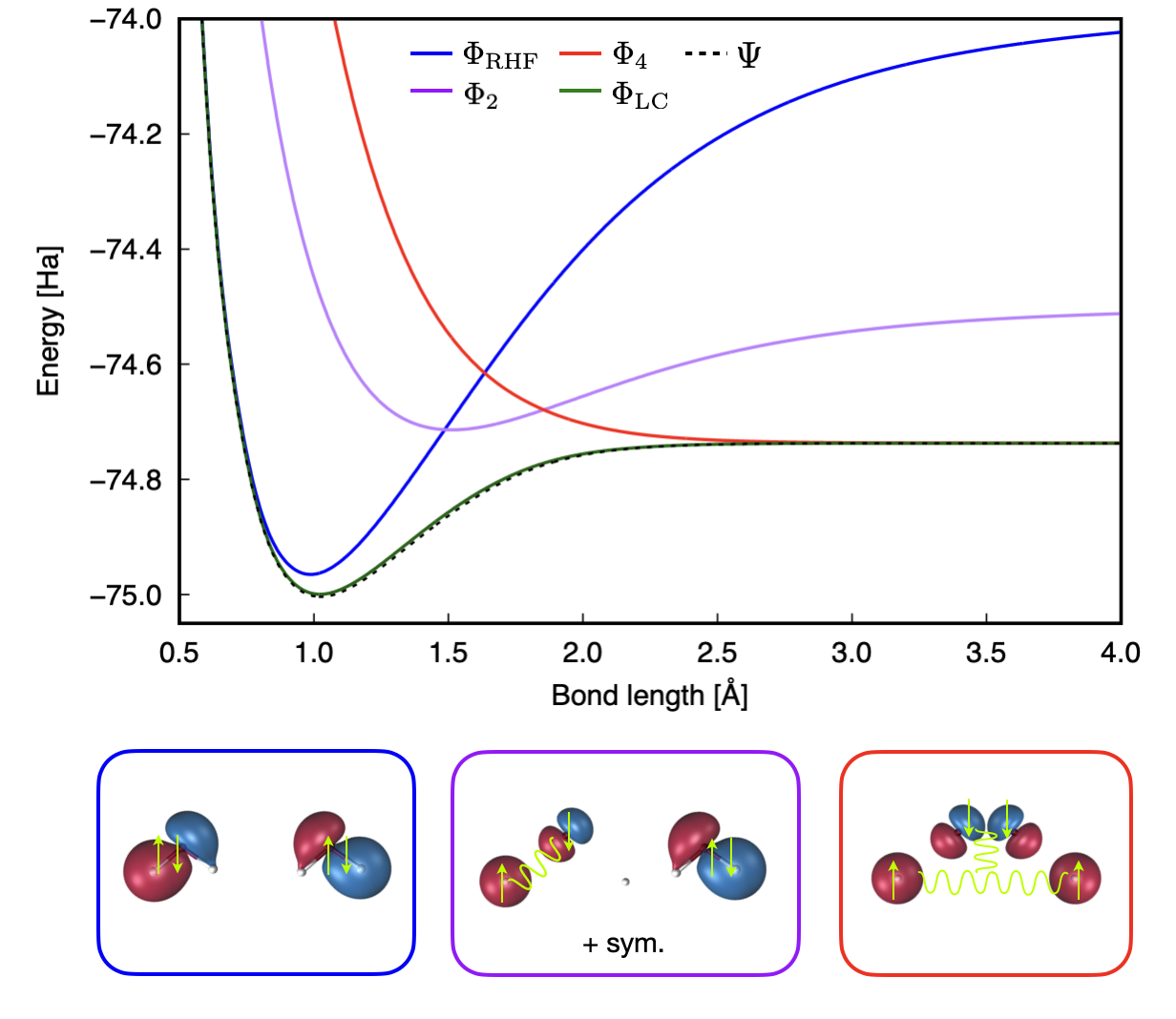}
		\label{sfig:h2o_energies}
	\end{subfigure}
	\begin{subfigure}{0.48\textwidth}
		\centering
		\includegraphics[width=\textwidth]{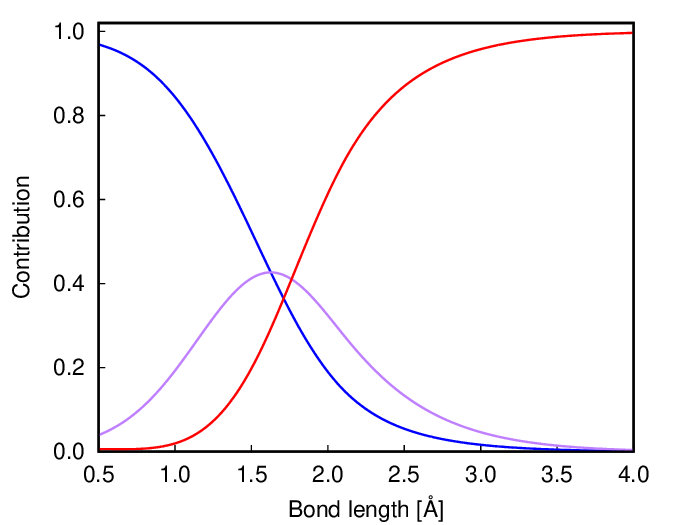}
		\label{sfig:h2o_coeffs}
	\end{subfigure}
	\centering
	\begin{subfigure}{0.48\textwidth}
		\centering
		\includegraphics[width=\textwidth]{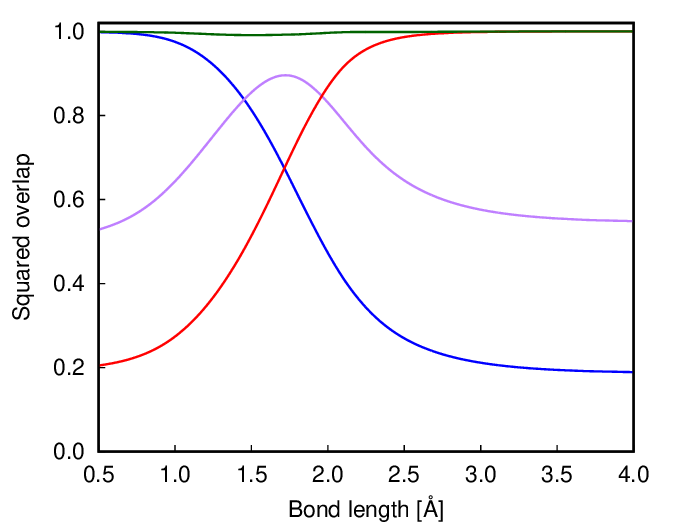}
		\label{sfig:h2o_overlaps}
	\end{subfigure}
	\hfill
	\centering
	\caption{Dissociation of H$_2$O in a STO-3G basis using generalised spin-coupled orbitals.}
	\label{fig:h2o_fullfig}
\end{figure}

The symmetric dissociation of \ce{H2O} is another challenging case for electronic structure theory, involving strong
correlation and spin-coupling among the electrons as the bonds break.\cite{Schaefer1980, Handy1981}
In contrast to \ce{N2}, the dissociation of \ce{H2O} involves the localisation of electrons onto more than two sites.

The MO description of water leads to the configuration $1\upa_1^2 2\upa_1^2 1\upb_2^2 3\upa_1^2 1\upb_1^2$, with term symbol ${}^1A_1$, 
where the $x$-axis points out of the plane of the molecule. The 
closed-shell orbitals form the state $\ket{ \mathcal{C}} = \ket{1\upa_1^2  2\upa_1^2 1\upb_1^2 }$, which contains
the $1\ups$, $2\ups$ and $2\upp_x$ orbitals on the oxygen atom, and the RHF wavefunction is
\begin{align}
\ket{\RHF} &= \ket{ \mathcal{C} } \ket { 1\upb_2^2 3\upa_1^2 } = \ket{ \mathcal{C} } \ket { \upsigma_L^2 \upsigma_R^2 }
\end{align}
where $\upsigma_L$ and $\upsigma_R$ are MOs localised on the left and right OH bonds.
The CSF for the dissociated state is that where the oxygen electrons from the two cleaved OH bonds
spin-couple to form a triplet oxygen state, due to the favourable exchange stabilisation, meaning 
that the two hydrogen electrons must also spin-couple to form a triplet and these triplets spin-couple to form the 
singlet molecular state:
\begin{align}
\ket{\Phi_4} &= \ket{ \mathcal{C} } \mathcal{O}^{4,1}_{00} ( \upo_L \upo_R \uph_L \uph_R )
\end{align}
where, in a minimal basis, the localised orbitals are obtained through simple rotation of the valence orbitals
\begin{align}
\nonumber
\upo_L &=  (\upsigma_{L} + \upsigma^\ast_{L} )/\sqrt 2\,, \qquad  \upo_R &=  (\upsigma_{R} + \upsigma^\ast_{R} )/\sqrt 2 \\
\nonumber
\uph_L &=  (\upsigma_{L} - \upsigma^\ast_{L} )/\sqrt 2\,, \qquad  \uph_R &=  (\upsigma_{R} - \upsigma^\ast_{R} )/\sqrt 2 \\
\nonumber
\upsigma_L &=  (1\upb_{2} + 3\upa_{1} )/\sqrt 2\,, \qquad  \upsigma_L^\ast &=  (2\upb_{2} + 4\upa_{1} )/\sqrt 2 \\
\upsigma_R &=  (1\upb_{2} - 3\upa_{1} )/\sqrt 2\,, \qquad  \upsigma_R^\ast &=  (2\upb_{2} - 4\upa_{1} )/\sqrt 2
\end{align}
The CSF that describes the intermediate configuration between the fully bonded and fully dissociated states
includes one broken OH bond. The molecular point group symmetry demands that
this state must be a symmetric superposition of the left and right bonded configurations, which results in 
\begin{align}
\ket{\Phi_2} &= \ket{ \mathcal{C} } \ket{\upsigma_L^2} \mathcal{O}^{2,1}_{00} ( \upo_R \uph_R ) 
+ \ket{ \mathcal{C} } \ket{\upsigma_R^2 } \mathcal{O}^{2,1}_{00} ( \upo_L h_L )
\end{align}
The relative contributions of each CSF to a variationally optimised linear combination follow the intuitive trend, 
transitioning from the RHF state at equilibrium to $\ket{\Phi_4}$ at dissociation (Figure~\ref{fig:h2o_fullfig}).
Remarkably, the electron correlation is almost entirely captured by the three CSFs at all bond lengths,
and the squared overlap with the exact wavefunction is at least 99\% everywhere.
At intermediate bond lengths all three CSFs have approximately equal contributions, which 
coincides with the region where traditional algorithms encounter the most difficulties.

\section{Hydrogen chains: \ce{H4}}\label{sec:h4}

The change in the electronic structure of hydrogen chains as the separation increases is another
challenging case for electronic structure theory. This simple system can be used to understand
conjugation in $\pi$ systems and is closely related to the Hubbard model for metal-to-insulator transitions in materials.\cite{Motta2017}

\begin{figure}
\centering
\includegraphics[width=\textwidth]{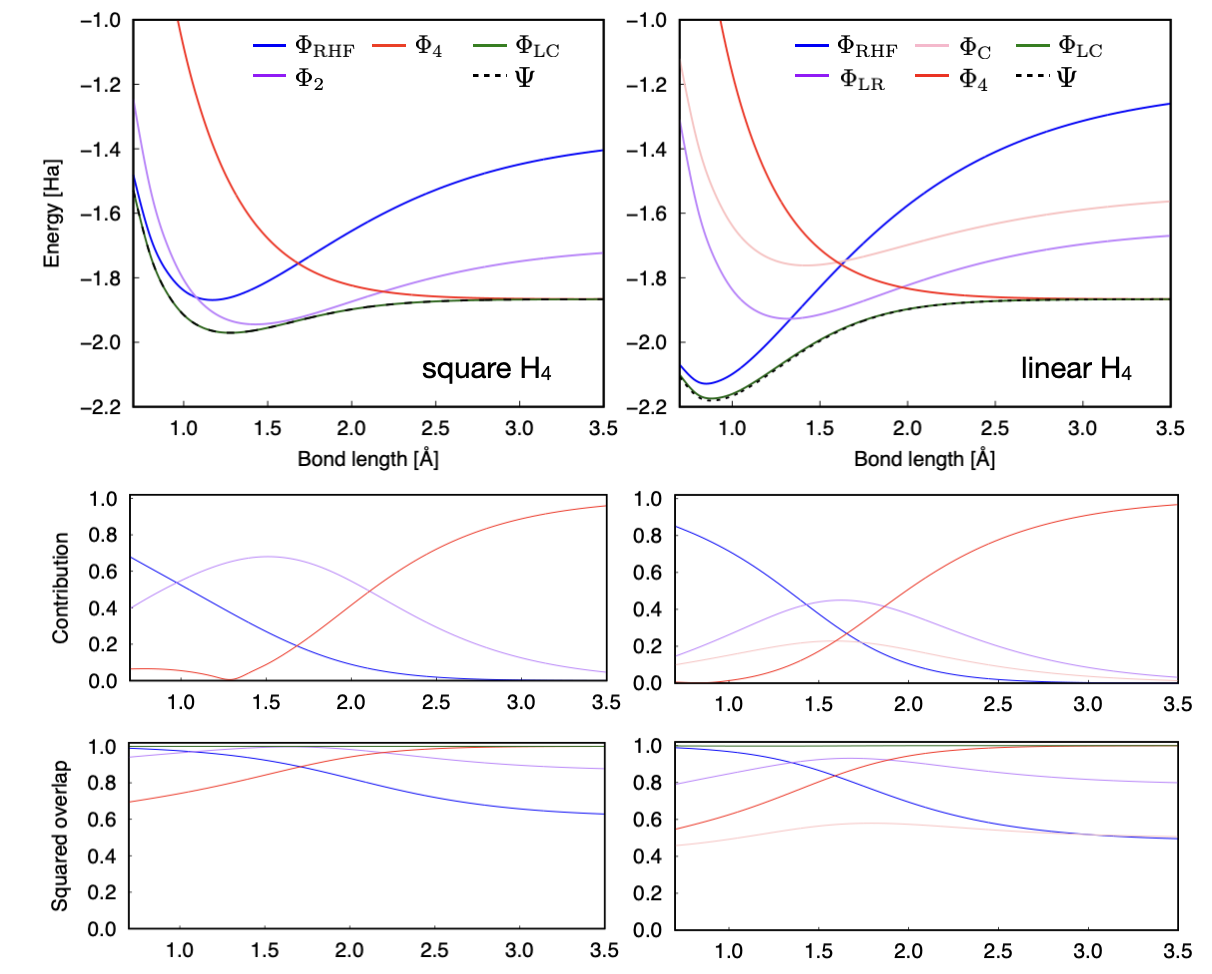}
\caption{Dissociation of square and linear \ce{H4} in a STO-3G basis.}
\label{fig:h4}
\end{figure}

We start with the cyclic \ce{H4} structure, which shares electronic structure characteristics with the $\uppi$ system of 
cyclobutadiene in a square geometry, and has a degenerate LUMO. In a minimal basis, the MOs are uniquely determined by symmetry and are the 
set $\upa_{1g},\upe_{u,x},\upe_{u,y},\upb_{2g}$. We consider the open-shell singlet
${}^1\mathrm{A_{1g}}$, which cannot be represented by a Slater determinant and is the CSF
\begin{align}
\ket{\Phi_0} &= \ket{ \upa_{1g}^2 \upe_{u,x} } + \ket{ \upa_{1g}^2 \upe_{u,y} }
\end{align}
where the $x$ and $y$ axes pass through the sides of the square.
Following our reasoning that the next most important configurations are those where spins localise,
we introduce bonding and antibonding orbitals localised on each of the four edges, left, right, top and bottom, 
through transformation of the valence orbitals
\begin{align}
\nonumber
\upsigma_L &=  (\upa_{1g} + \upe_{u,x} )/\sqrt 2\,, \qquad  \upsigma_L^\ast &=  (\upb_{2g} + \upe_{u,x} )/\sqrt 2 \\
\nonumber
\upsigma_R &=  (\upa_{1g} - \upe_{u,x} )/\sqrt 2\,, \qquad  \upsigma_R^\ast &=  (\upb_{2g} - \upe_{u,x} )/\sqrt 2 \\
\nonumber
\upsigma_T &=  (\upa_{1g} + \upe_{u,y} )/\sqrt 2\,, \qquad  \upsigma_T^\ast &=  (\upb_{2g} + \upe_{u,y} )/\sqrt 2 \\
\upsigma_B &=  (\upa_{1g} - \upe_{u,y} )/\sqrt 2\,, \qquad  \upsigma_B^\ast &=  (\upb_{2g} - \upe_{u,y} )/\sqrt 2 
\end{align}
The atomic $1\ups$ orbitals are recovered by taking linear combinations of bonding and antibonding pairs. 
The CSF that describes the valence state with partial localised bonding is
\begin{align}
\ket{\Phi_2} &= \ket{ \upsigma_L^2 } \mathcal{O}^{2,1}_{00} (\ups_3,\ups_4) + \ket{ \upsigma_R^2 } \mathcal{O}^{2,1}_{00} (\ups_1,\ups_2) \nonumber\\
             &\,\, + \ket{ \upsigma_T^2 } \mathcal{O}^{2,1}_{00} (\ups_1,\ups_4) + \ket{ \upsigma_B^2 } \mathcal{O}^{2,1}_{00} (\ups_2,\ups_3)
\end{align}
and the CSF for the fully open-shell state is
\begin{align}
\ket{\Phi_4} &= \mathcal{O}^{4,2}_{00} (\ups_1,\ups_2,\ups_3,\ups_4) + \mathcal{O}^{4,2}_{00} (\ups_2,\ups_3,\ups_4,\ups_1)
\end{align}
where $\mathcal{O}^{4,2}_{00}$ is the antisymmetrised product of two open-shell singlets (Eq \ref{eqn:vcsf_n4_2}).
In Figure \ref{fig:h4} we present the binding curves for the symmetric dissociation of H$_4$. These three valence states
are sufficient to reproduce the exact binding curve and represent all of the correlation processes involved.

We now turn to the linear H$_4$, which shares electronic characteristics with the $\uppi$ system of butadiene.
The valence states for linear H$_4$ are the same as those of the cyclic structure, but where left-right and top-bottom
symmetry was present in the square geometry it is now lost, becoming the left-right and centre-edge. The left-right
and centre contributions to $\ket{\Phi_2}$ must be included separately, and the edge contributions
become unimportant since they are non-bonding. At the dissociation limit, all $2^4$ spin states with one electron on each atom
are degenerate, but the degeneracy is lifted by the kinetic energy of delocalisation at finite separations. The state 
$\ket{\Phi_4}$ is the linear combination of $S=0$ CSFs with maximal nearest neighbour delocalisation energy, and
equivalently, minimal nearest neighbour spin coupling 
$\langle \Phi \vert \hat S_i\cdot \hat S_j \vert \Phi \rangle$ through the isomorphism with the Heisenberg--Dirac spin 
Hamiltonian.\cite{Dirac1929a,Anderson1963, ClevelandMedina1975}

\section{Quantum computation using spin-coupled states}\label{sec:quantum_algos}

In digital quantum computing, the many-electron wavefunction is encoded by mapping it to a quantum superposition of the qubit
states of the quantum device. Quantum algorithms involve preparing the initial qubit state for a known reference wavefunction $\ket{\Phi_\text{ref}}$,
and either operating on it to drive it to ground state of the Hamiltonian $\ket{\Psi}$, 
for example in quantum imaginary time-evolution\cite{Motta2019}
and adiabatic state preparation,\cite{Aspuru-Guzik2005} or time-evolving it to extract the ground state energy through probabilistic projective measurements, for example 
in quantum phase estimation (QPE).\cite{Kitaev1995, Aspuru-Guzik2005, Abrams1999, Nielsen2010}
The cost of these approaches depends critically on the overlap of the initial reference state with the exact ground state wavefunction.\cite{Lee2023}

In QPE, for example, the probability of measuring the ground state energy 
is determined by the squared overlap $\gamma^2 = \abs{\braket{\Phi_\text{ref}|\Psi}}^2$. Consequently, 
$\mathcal{O}(\abs{\gamma}^{-2})$ measurements are required to obtain the ground state energy reliably.
This overlap problem remains significant despite the tremendous progress over the past two decades, which has reduced the quantum circuit cost (number of operations) required to implement time-evolution to scale approximately quadratically with the number of orbitals,\cite{Low2019, Berry2019, Burg2021, Lee2021} and has seen the emergence of 
post-QPE approaches requiring $\mathcal{O}(\abs{\gamma}^{-1})$ measurements.\cite{Lin2020a, Lin2022, Dong2022} 
.

Most implementations of quantum algorithms employ the Hartree--Fock state as the reference.\cite{McArdle2020}
However, since the overlap between the Hartree--Fock determinant and the exact ground state decays exponentially with the number of 
open-shell electrons, the performance of subsequent quantum algorithms for strongly-correlated systems rapidly deteriorates.\cite{Lee2023}
Several works on concrete resource estimates for QPE-based electronic structure for challenging systems predict runtimes for a single-point energy calculation in the order of several days, despite generous assumptions for the quantum hardware requirements.\cite{Reiher2017, Burg2021, Goings2022} It is clear that reductions in the overall runtime of QPE, as enabled through improved reference states,
are of paramount importance to enable practical quantum computation of electronic structure.

Our spin-coupled formalism provides a systematic approach to constructing 
reference wavefunctions with good overlaps with the exact state, and therefore has the potential to unlock
the power of quantum algorithms for challenging chemical problems.\cite{Reiher2017,Sugisaki2019,Lee2021} 
Our approach is ideally suited to quantum computing. Although the
number of determinants in each CSF is equal to the number of fully open-shell determinants with $M_s = 0$, which grows combinatorially with the size of the system as ${N\choose N/2}$,\cite{Paldus1974} the number of distinct coefficients only scales as $\mathcal{O}(N)$.
Their prescribed entanglement structure means that they can be efficiently prepared on a quantum device, with a cost negligible to that of quantum algorithms.\cite{Marti-Dafcik2024b} 
Similarly, the fact that each contributing valence state uses different sets of orthonormal orbitals, and that the 
valence states are not mutually orthogonal, poses no problem for quantum state preparation.\cite{Marti-Dafcik2024b}
Using our reference wavefunctions as initial states in QPE directly 
reduces the overall runtime. In the simple case of \ce{N2}, for example, $\vert\braket{\Phi_\text{RHF}|\Psi}\vert^2\sim0.06$ at elongated bonds, 
but $\vert\braket{\Phi_\text{LC}|\Psi}\vert^2\sim1$ (Fig.~\ref{fig:n2_csf_fid}), which translates to a sixteen-fold speed-up in QPE execution.

Quantitative quantum chemical predictions\cite{Helgaker2008} require the simultaneous 
treatment of strongly interacting states, dynamic correlation and orbital relaxation. 
Where classical methods fail due to problems of complexity,\cite{Evangelista2018} the quantum approach extends straightforwardly.
The orbitals for each configuration can be optimised in a large basis on a classical computer using
RASSCF techniques at only mean-field cost. The linear coefficients of the accurate reference state 
can be computed in the same manner as quantum subspace algorithms,\cite{Stair2020a, Klymko2022} where each pair of CSFs 
is prepared on the quantum computer and overlap and Hamiltonian matrix elements obtained through measurement.
Since the reference state directly encodes a large part of the entanglement structure of 
electronic ground states, a high-quality ground state energy can be extracted efficiently through QPE.

\section{Conclusions and outlook}

Despite the fact that the chemical bond is at the very heart of chemical theory, an accurate representation 
of the process of bond formation has presented a severe challenge to quantum chemical methods. 
Modern approaches\cite{Andersson1990,Andersson1992,Bartlett2007,Szalay2012,Evangelista2018}
are based on complete expansions within a set of active MOs\cite{Roos1980a} and suffer from high computational and operational complexity,
which has encouraged the widespread erroneous belief that the wavefunction is inherently complex. 
Using the spin-coupled molecular orbital theory presented in this article, we have shown that the wavefunction
is in fact highly structured and that bond formation can be represented through a handful of spin-coupled states, each with
well-defined spin quantum numbers and physical characteristics.

Our spin-coupled theory exposes a simple interpretation of the chemical bond. The character of the bond at 
equilibrium bond lengths is predominantly that predicted by MO theory, but electron crowding is relieved by
diradical character, where electrons localise on opposite atoms, leading to a reduction in formal bond order and 
bond elongation. As a bond breaks, the diradical character increases and the MO character reduces. 
For multiple bonds, the radical states are characterised by high-spin coupling of the electrons localised on each atom.
The $\upsigma$ and $\uppi$ bonds break at different bond lengths, with the $\uppi$ bonds typically breaking first, and the bond
order reduces sequentially as each pair of electrons in the bonds localise on opposite atoms.
Applying this simple theory recovers 90\% of the wavefunction resulting from a complete active space calculation, 
and our simple chemical model therefore has a direct connection to sophisticated wavefunction expansions. 

Our model invokes different orbitals for each configuration, and the resulting states are not orthogonal.
Moreover, the spin-coupled configurations correspond to different genealogical spin-coupling schemes, with 
different orbital orderings. The highly compact representation of the wavefunction afforded by our approach is lost
when either a single orbital basis, or single CSF basis is used to represent all contributing states. 
For the archetypical strongly correlated systems considered in this work, the apparent complexity of the 
wavefunction in traditional approaches arises from the insistence on using a single basis of orthonormal
orbitals and a single CSF basis to represent all configurations, so that 
the Hamiltonian matrix evaluation is straightforward.\cite{HelgakerBook,Paldus1974,Shavitt1977} 
While matrix elements between nonorthogonal Slater determinants can be readily computed,\cite{MayerBook,Burton2021c,Burton2022c} 
building the determinant expansion of each CSF has exponential cost using traditional algorithms.
To exploit the compact structure revealed by our theory, it is necessary to use methods
that can handle non-orthogonal CSF states. 

Quantum computing has the potential to overcome many of the remaining challenges in quantum chemistry.
Provided that a high-quality reference state can be efficiently parametrised as a quantum circuit,
algorithms such as quantum phase estimation and quantum eigenstate filtering can efficiently extract the exact
ground state within a given basis set. Our spin-coupled molecular orbital theory provides a framework for
constructing highly entangled reference states for strongly correlated molecular ground states using only 
a basic understanding of a chemical process, a small number of variational parameters, and at the cost of a mean-field calculation.
Since the entanglement structure of each configuration is defined through spin and symmetry constraints, they can be
prepared and transformed on a quantum register with low-depth quantum circuits, providing a practical
route to using quantum algorithms to address challenging problems in quantum chemistry.

\acknowledgments
D.M.D. thanks Alexander Gunasekera for providing helpful suggestions and discussing several aspects of this work, and acknowledges financial support by the EPSRC Hub in Quantum Computing and Simulation (EP/T001062/1). H.G.A.B. acknowledges financial support from New College, Oxford (Astor Junior Research Fellowship) and Downing College, Cambridge (Kim and Julianna Silverman Research Fellowship).

\appendix



\section{Spin eigenfunctions}\label{sec:spin_eigenfunctions}

A CSF $\mathcal{O}_{SM}^{N,i}(\phi_1 \dots \phi_N )$ has $N$ electrons in spatial orbitals $\{\phi\}$
and is a spin eigenfunction with quantum numbers $S$ and $M$. Each CSF is a linear
combination of determinants, here represented through the $\alpha$ or $\beta$ occupation of the orbitals $\phi$.

	\begin{equation}
		\ket{\mathcal{O}_{0, 0}^{2, 1}} = \frac{1}{\sqrt{2}}(\ket{\al\be} - \ket{\be \al}),
		\label{eqn:vcsf_n2}
	\end{equation}
	\begin{equation}
                \begin{split}
                        \ket{\mathcal{O}_{1, 1}^{2, 1}}=
                        \ket{\alpha\alpha},
                \end{split}
                \label{eqn:tripcsf_n2_1}
        \end{equation}
	\begin{equation}
		\begin{split}
			\ket{\mathcal{O}_{0, 0}^{4, 1}}=
			\frac{1}{\sqrt{3}}
			(\ket{\alpha\alpha\beta\beta}+\ket{\beta\beta\alpha\alpha})  \\
			- \frac{1}{2\sqrt{3}}(\ket{\alpha\beta\alpha\beta}+\ket{\beta\alpha\beta\alpha}
			+\ket{\alpha\beta\beta\alpha}
			+\ket{\beta\alpha\alpha\beta}),
		\end{split}
		\label{eqn:vcsf_n4}
	\end{equation}
        \begin{equation}
                \begin{split}
                        \ket{\mathcal{O}_{0, 0}^{4, 2}}=
                        \frac{1}{2}
                        (\ket{\alpha\beta\alpha\beta}
                        -\ket{\alpha\beta\beta\alpha}
                        -\ket{\beta\alpha\alpha\beta}
                        +\ket{\beta\alpha\beta\alpha}),
                \end{split}
                \label{eqn:vcsf_n4_2}
        \end{equation}
	\begin{equation}
                \begin{split}
                        \ket{\mathcal{O}_{1, 1}^{4, 1}}=
                        \frac{1}{2}
                        (\ket{\alpha\alpha\alpha\beta}
                        -\ket{\alpha\alpha\beta\alpha}
                        -\ket{\alpha\beta\alpha\alpha}
                        +\ket{\beta\alpha\alpha\alpha})
                \end{split}
                \label{eqn:tripcsf_n4_1}
        \end{equation}
	\begin{equation}
		\begin{split}
			\ket{\mathcal{O}_{0, 0}^{6, 1}} =
			\frac{1}{2}(\ket{\alpha\alpha\alpha\beta\beta\beta}) - \ket{\beta\beta\beta\alpha\alpha\alpha})\\
			+\frac{1}{6}(\ket{\alpha\beta\beta\alpha\alpha\beta}
			+ \ket{\alpha\beta\beta\alpha\beta\alpha}
			+ \ket{\alpha\beta\beta\beta\alpha\alpha}
			+ \ket{\beta\beta\alpha\beta\alpha\alpha}\\
			+ \ket{\beta\beta\alpha\alpha\beta\alpha}
			+ \ket{\beta\beta\alpha\alpha\alpha\beta}
			+ \ket{\beta\alpha\beta\beta\alpha\alpha}
			+ \ket{\beta\alpha\beta\alpha\beta\alpha}\\
			+ \ket{\beta\alpha\beta\alpha\alpha\beta}
			- \ket{\beta\alpha\alpha\beta\beta\alpha}
			- \ket{\beta\alpha\alpha\beta\alpha\beta}
			- \ket{\beta\alpha\alpha\alpha\beta\beta}\\
			- \ket{\alpha\beta\alpha\beta\beta\alpha}
			- \ket{\alpha\beta\alpha\beta\alpha\beta}
			- \ket{\alpha\beta\alpha\alpha\beta\beta}
			- \ket{\alpha\alpha\beta\beta\beta\alpha}\\
			- \ket{\alpha\alpha\beta\beta\alpha\beta}
			- \ket{\alpha\alpha\beta\alpha\beta\beta}
			),\end{split}
		\label{eqn:vcsf_n6}
	\end{equation}


\section{Computational details}
We used PySCF to obtain the integrals and compute the RHF and FCI solution,\cite{Sun2018, Sun2020} and an in-house Python code for all other tasks, including: generating the spin eigenfunctions from the Clebsch-Gordan coupling coefficients, performing basis transformations, computing matrix elements and wavefunction overlaps, and solving the generalized eigenvalue problem. The visualization of the molecular orbitals was done using VMD.\cite{VMD}

 \bibliographystyle{science}

\bibliography{export,hughlib}

\end{document}